\documentclass{article}

\usepackage{arxiv}

\usepackage[utf8]{inputenc} 
\usepackage[T1]{fontenc}    
\usepackage{hyperref}       
\usepackage{url}            
\usepackage{booktabs}       
\usepackage{amsfonts}       
\usepackage{nicefrac}       
\usepackage{microtype}      
\usepackage{lipsum}		
\usepackage{graphicx}
\usepackage[square,sort&compress,numbers]{natbib}
\usepackage{doi}
\usepackage{braket}
\usepackage{tabularx}
\usepackage{amsmath}
\usepackage{mathtools}

\title{Synchronized Bimodal Amplitude Patterns in Heterogeneous Oscillatory Media — Experiment and Theory}

\usepackage{authblk}

\author[1]{Nicolas Thomé}
\author[1]{Yukiteru Murakami}
\author[1\thanks{\tt{Krischer@tum.de}}]{Katharina Krischer}

\affil[1]{School of Natural Sciences, Physics Department, Nonequilibrium Chemical Physics, Technische Universität München, James-Franck-Str. 1, D-85748 Garching, Germany}

\hypersetup{
pdftitle={A template for the arxiv style},
pdfsubject={},
pdfauthor={David S.~Hippocampus, Elias D.~Striatum},
pdfkeywords={First keyword, Second keyword, More},
}

\begin{document}
\maketitle

\begin{abstract}
We study an intricate mechanism of pattern formation in globally coupled heterogeneous oscillatory media. In anodic electrochemical etching of silicon, the electrode surface splits into two amplitude–phase regions, while all oscillators remain frequency-locked.
Additionally, the relative ratio of the pattern can be tuned via a coupling term. We introduce a heterogeneous, complex Ginzburg–Landau equation with global coupling to reproduce these patterns and study their formation. Neglecting diffusion shows that frequency entrainment arises from amplitude adaptation. Diffusion, on the other hand, enforces the selection of a unique cluster ratio. In quantitative agreement with simulations, a center-manifold reduction yields a Lyapunov functional that predicts the selected ratio. Both of these results establish a theoretical framework that connects experiment and theory. Moreover, they show how heterogeneity and diffusion convert degenerate cluster dynamics into robust pattern selection.
\end{abstract}


\section{\label{sec:Intro}Introduction}
Oscillations are a fundamental dynamical state that determines the functioning of many natural and human-made systems. Macroscopic oscillations are often observed in systems comprising many interacting microscopic oscillatory units and result from synchronization, where the rhythms of the individual oscillating entities are adjusted to a common frequency~\cite{Pikovsky.2010}. This underscores the importance of understanding the emergence of collective behavior in ensembles of coupled oscillators.\ 

Key synchronization features are captured by the paradigmatic Kuramoto model that reduces the dynamics of an oscillator to the dynamics of its phase~\cite{Kuramoto.1975, Haken.1984, Strogatz.2000}. In particular, the model captures the transition from an incoherent to a partially synchronized state in an ensemble of globally coupled oscillators with a unimodal frequency distribution as the coupling strength increases. The theoretically predicted second-order-type transition could be validated experimentally with an array of globally coupled electrochemical oscillators~\cite{Kiss.2002}. Other diverse experimental realizations range from the synchronization of generic mechanical metronomes~\cite{Pantaleone.2002} to the synchronization of oscillators as complex as Josephson junctions~\cite{Wiesenfeld.1998}.\

Because of its simplicity and universality, the Kuramoto model has been extended in various directions. In particular, various coupling topologies, coupling functions, types of frequency distributions, or additional external fields have been employed. Comprehensive reviews summarize these extensions and some of their applications in different disciplines~\cite{Acebron.2005,Gupta.2014}.

However, there exist synchronization patterns that exhibit pronounced amplitude variations~\cite{Wang.2000, Schmidt.2014, Patzauer.2021}. This typically happens when the coupling strength is large. Clearly, phase models fail in this coupling regime. Instead, modeling approaches often use networks of Stuart-Landau (SL) oscillators, generic oscillators close to a supercritical Hopf bifurcation~\cite{Guckenheimer.1983}. Compared to phase models, the complexity of systems of coupled SL oscillators is significantly greater, and our understanding of their dynamics remains in its infancy. Most investigations consider identical and globally coupled SL oscillators with a focus on identifying qualitatively new types of collective states. Among the described collective dynamics are amplitude cluster solutions, chimera states, and incoherent solutions that adapt amplitudes off the unit circle~\cite{Nakagawa.1993, Nakagawa.1994, Chabanol.1997, Sethia.2014, Clusella.2019}. Other studies investigate the multiplicity of the solutions which result from the inherent full permutation symmetry of systems of identical oscillators~\cite{Golubitsky.2002,Kemeth.2019, Kemeth.2020, Ku.2015, Thome.2025}. In contrast, the literature on the impact of heterogeneous frequency distributions in an ensemble of SL oscillators is still sparse and focused on a few parameter regimes~\cite{Aronson.1990, Matthews.1991, Monte.2002, Millan.2025}. \

So far, we have considered ensembles of discrete oscillators. In many physical systems, however, the situation is more involved. This applies in particular to globally coupled oscillating continuous media, where, in addition to the global coupling, oscillators interact with each other due to transport processes such as diffusion. Prominent examples are systems that are electrically controlled, such as gas discharge tubes~\cite{Ammelt.1997} and electrode reactions~\cite{Krischer.2002}, but also the light-sensitive Belousov-Zhabotinsky reaction~\cite{Vanag.2000, Nicola.2006}, or the heterogeneous CO oxidation on Pt single crystal catalysts~\cite{Falcke.1995}; observed patterns include cluster formation, standing waves, and stripe pattern. The theoretical description of these systems was achieved with the complex Ginzburg-Landau equation (CGLE) extended by a global coupling term~\cite{Falcke.1995, Yochelis.2004}.\

Here, we introduce a type of amplitude pattern that forms in experiments involving an oscillatory continuous medium with a strong global and a weak diffusional coupling. The pattern is captured by an extended CGLE with small heterogeneities. However, since it is dominated by global coupling, the mechanism by which it forms can be elucidated using a heterogeneous system of globally coupled SL oscillators. Furthermore, we demonstrate that the solution inherits its ordering principle in parameter space from the cluster singularity~\cite{Kemeth.2019} found in globally coupled identical SL oscillators. In addition, treating diffusional coupling as a small perturbation, we can derive a Lyapunov functional and demonstrate that in the presence of a small diffusion, the system selects a particular state among the large amount of coexisting stable patterns in discrete globally coupled oscillators. We thus combine the various approaches mentioned above to understand collective oscillations, enabling us to demonstrate how disturbances in real systems influence their dynamics. 

Experimentally, we study the anodic electrochemical dissolution of silicon, a wet-chemical preparation method for obtaining smooth Si surfaces~\cite{XiaogeGregoryZhang.2004}. When silicon is immersed in fluoride-containing electrolyte under anodic bias, the dissolution process does not proceed steadily, but instead generates self-sustaining oscillations in oxide thickness and current~\cite{Turner.1958, Blackwood.1992}. These oscillations indicate nonlinear feedback loops between interfacial reactions, charge transport, and oxide growth. When using n-doped Si wafers, patterns in the oxide layer thickness are observed where typically the amplitudes of different oscillating regions vary significantly ~\cite{Schonleber.2014,Schmidt.2014}. In this article, we introduce a particular experimentally obtained collective state: \emph{a bimodal amplitude solution}, i.e., a pattern characterized by two regions, in which the amplitudes of the oscillators are centered around one of the two modes within a region. At the same time, the entire electrode oscillates at a uniform frequency, i.e., it exhibits complete frequency synchronization. Interestingly, the ratio between the two amplitude areas could be continuously tuned via a global coupling parameter.

The paper is structured as follows. In Section~\ref{sec:ECSE}, we describe the experimental setup and introduce the observed bimodal amplitude patterns.
To reproduce these experimental states numerically, in Section~\ref{sec:CGLEMFsec}, we employ a system of globally coupled heterogeneous Stuart-Landau oscillators with diffusion. 
In Section~\ref{sec:Discussion}, we further investigate the bimodal amplitude distribution and examine the impact of heterogeneity and diffusion on a system of globally coupled SL oscillators.
Section~\ref{subsec:HChetSL} ignores diffusion and analyzes how amplitude dynamics enforce frequency locking using self-consistency tools. Section~\ref{subsec:PatSelec} develops a Lyapunov functional that predicts which one of the coexisting patterns is selected. We conclude in Section~\ref{sec:Conclusion} with a discussion of the broader implications for synchronization and pattern formation in nonlinear media. 

\section{\label{sec:ECSE}Experiment with Electrochemical Silicon Dissolution System}

The impact of global coupling on the formation of amplitude patterns was investigated experimentally using an electrochemical oscillator, silicon electrooxidation in a fluoride-containing electrolyte. For an overview of the anodic behavior of Si electrodes, see chapter 5 in~\cite{XiaogeGregoryZhang.2004}, details of the oscillation mechanism can be found in~\cite{Murakami.}. 
The Si oscillator is known to exhibit amplitude clusters at certain parameter values~\cite{Patzauer.2021}.
In this study, we control the degree of global coupling and demonstrate how it influences the characteristics of the amplitude patterns.

\subsection{Setup}
The experimental setup consists of an electrochemical cell, an illumination setup, and an ellipsometric imaging device.
The working electrode is a single-crystalline n-type silicon wafer (111) with a resistivity of $1-10 \;\rm  \Omega cm$ cut into $7 \times 7 \,\mathrm{mm}^2$ squares.
A $200\,\rm nm$ aluminium film is deposited as a back contact and annealed at $250\,^\circ \rm C$ for 15 minutes.
The surface is plasma-oxidized to remove organic contamination and passivate the surface.
The silicon sample is mounted on a custom-made PTFE (polytetrafluoroethylene) holder.
Conductive silver paste is used for electrical contact, and silicone rubber is used for sealing. A platinum wire and a saturated $\rm Hg|Hg_{2}SO_4$ electrode served as counter and reference electrodes, respectively.
The electrolyte is an aqueous solution of $500 \,\rm ml$ containing $60 \,\rm mM$ $\rm NH_4F$ and $142 \,\rm mM$ $\rm H_2SO_4$, which yields $\rm pH = 1$ at room temperature.
The electrolyte is deaerated by bubbling with argon for 20 minutes before the first measurement.
During the measurements, the electrolyte is stirred with a magnetic stirrer.
A VSP potentiostat (BioLogic) is used for potentiostatic control.
A variable resistor is connected in series to the working electrode.

Illumination is performed with a laser (He--Ne, $632\,\rm nm$) to generate valence band holes necessary for the electrochemical reaction.
The illumination intensity is adjusted to $0.88 \,\rm mW/cm^2$ so amplitude clusters can be observed~\cite{Patzauer.2021} using an electrically controlled gray filter.

During the electrochemical measurements, an ellipsometric imaging device monitors the spatiotemporal evolution of the silicon oxide layer formed at the interface between the working electrode and the electrolyte.
The recorded signal is the light intensity reflected at the interface.
The signal varies with the optical path length, including changes in the oxide's thickness and the dielectric constant at the interface.
The details of the ellipsometric setup are described in Ref.~\cite{Miethe.2009}. 

The degree of global coupling is controlled by the value of the external resistor, $R_{\rm ext}$, which is connected in series with the electrode.
The potential drop across the resistor depends on the total current through the electrode. It modulates the effective potential drop, $U_{\rm eff}$, which in turn determines the local currents, $j_{\rm loc}(\vec{x})$.
If we denote the applied potential as $U_{\rm app}$,  $U_{\rm eff}$ is given by
\begin{equation}
\label{eq:resitanceGC}
U_{\rm eff} = U_{\rm app} -  R_{\rm ext} \int_{\rm electrode} j_{\rm loc}(\vec{x}) dS,
\end{equation}
where the integration runs over the electrode area, $A$. The electrode area was $A = 14.52 \,\rm mm^2$ in all measurements.
In this system, the local electrochemical reaction is mediated by holes in the valence band, and the formation of patterns is known to be governed by the horizontal motion of the holes on the surface of the electrode, resulting from a reaction–diffusion–migration equation with an additional nonlinear, nonlocal term (see Eq. (5) in the Supplementary Material of Ref.~\cite{Patzauer.2021}). 
Therefore, the local density of the holes, controlled by $U_{\rm eff}$, is globally coupled through the external resistor as indicated in equation~\eqref{eq:resitanceGC}.
The higher the resistance, the stronger the global coupling.
We use $R = R_{\rm ext}\cdot A \,\rm [k\Omega cm^2]$ to quantify the global coupling intensity hereafter.
To keep the local reaction rate practically independent of the strength of the global coupling, we adjusted $U_{\rm app}$ such that $U_{\rm eff}$ was always in the same range.
More precisely, $R_{\rm ext}$ was changed in $3 \,\rm k\Omega$ increments and at each step $U_{\rm app}$ was increased by $0.13 \,\rm V$.
In each step, the parameter was kept constant for 400~s after the initial transient period.
The following table shows the pairs of investigated parameters $R$ and $U_{\rm app}$.

\begin{table}[htbp]
\centering
\begin{tabularx}{0.9\textwidth}
{ 
  | l
  | *{9}{>{\raggedright\arraybackslash}X} |
  }
 \hline
 $R \rm\,[k\Omega cm^2]$ &0.58 &1.02 &1.46 &1.90 &2.33 &2.77 &3.21 &3.65 &4.09\\
 \hline
 $U_\mathrm{app}\,[\mathrm{V~vs~MSE}]$ &3.24  & 3.37  &3.50 &3.63 &3.76 & 3.89 & 4.02 & 4.15& 4.28\\
\hline
\end{tabularx}
\caption{Parameter pairs of the resistance $R$ and the applied potential $U_{\rm app}$ used in the measurements.}
\label{tab:Ex_parameters}
\end{table}

\subsection{Data analysis}
We now outline the two procedures employed in our data analysis.
\paragraph{Hilbert Transformation}
The recorded images from the ellipsometric imaging device were first filtered with a Gaussian filter. The resulting data $\xi(\vec{x},t)_{\rm raw}$ were then normalized in the following way to account for inherent, small differences in the local average signal intensity:
\begin{equation}
\xi(\vec{x},t)= \left(\xi(\vec{x},t)_{\rm raw}- \overline{\xi(\vec{x},t)_{\rm raw}} \right) \cdot \frac{\braket{\overline{\xi(\vec{x},t)_{\rm raw}}}}{\overline{\xi(\vec{x},t)_{\rm raw}}}.
\end{equation}
Here, the overlines and cornered brackets represent the time and spatial averages, respectively.
Next, we performed a Hilbert transform of $\xi(\vec{x},t)$, resulting in $\xi_{\rm H}(\vec{x},t)$, which allows us to define the analytical signal $\zeta(\vec{x},t)$ that is finally expressed in terms of the amplitude $A(\vec{x}, t)$ and phase $\phi(\vec{x}, t)$ of the local oscillators
\begin{equation}
\zeta(\vec{x},t) = \xi(\vec{x},t) +{\rm i} \xi_{\rm H}(\vec{x},t) = A(\vec{x}, t){\rm e}^{{\rm i}\phi(\vec{x}, t)}.
\end{equation}
The local frequency of the oscillations, $\omega(\vec{x})$ is obtained from the slope of a linear fitting of the phase $\phi(\vec{x},t)$ over a certain time interval.
\paragraph{Kernel Density Estimator}
The amplitudes $A(\vec{x},t_0)$ from the Hilbert transformation at time step $t_0$ follow some unknown distribution. To obtain a smooth estimate, we applied a kernel density estimation (KDE):
\begin{equation}\label{eq:KDE}
\begin{aligned}
    \hat{f}(A)=\frac{1}{Nh}\sum_i^N K\left(\frac{A-A_i}{h}\right),
\end{aligned}
\end{equation}
where the kernel $K$ is a Gaussian. The bandwidth $h$ controls the smoothing; its default value was chosen according to Silverman's rule of thumb~\cite{Sheather.2004}.

\begin{figure*}[htbp]
\centering
\includegraphics[width=0.8\textwidth]{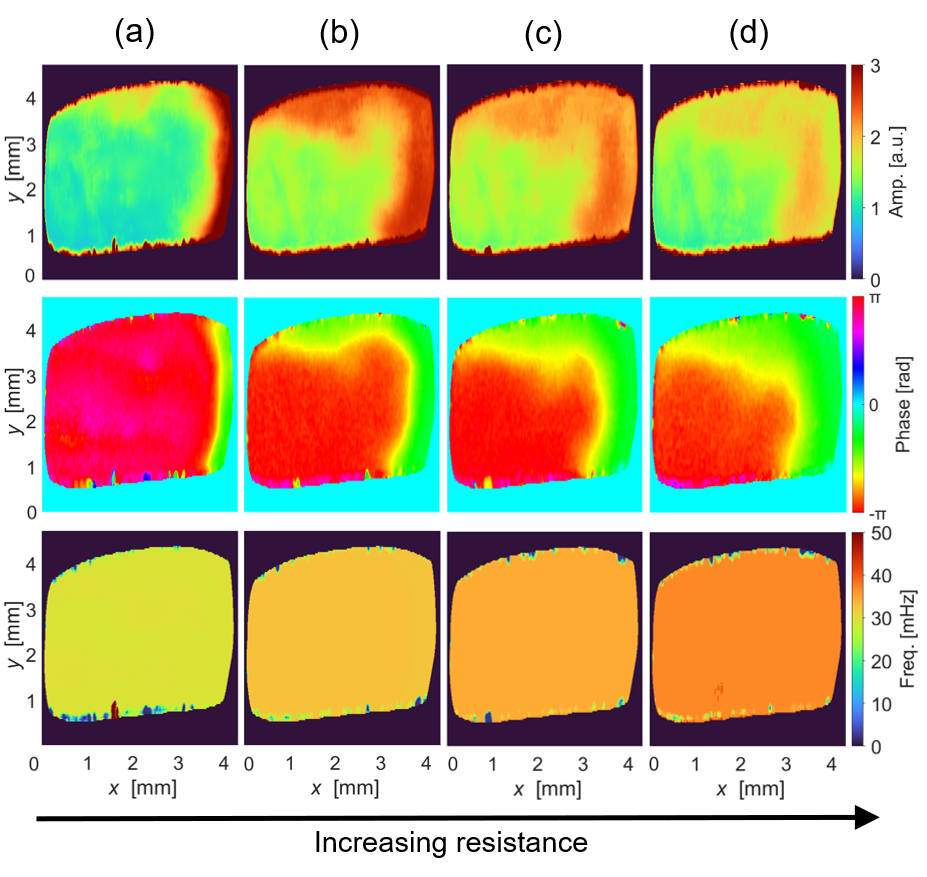}
\caption{
Experimentally measured distribution of time-averaged amplitude (first row), snapshots of phase distribution (second row), and time-averaged frequency distribution (third row) on an electrode surface during anodic Si electrodissolution. From (a) to (d) the external resistance was increased: (a) $R=0.58 \rm \,k\Omega cm^2$, (b) $R=1.46 \rm \,k\Omega cm^2$, (c) $R=2.33 \rm \,k\Omega cm^2$, and (d) $R=3.65 \rm \,k\Omega cm^2$.
The resistance was increased in steps, as shown in Table \ref{tab:Ex_parameters}. 
At each step, we waited 400~s after the initial transient state.
}
\label{fig:Ex_3x4}
\end{figure*}
\subsection{Results}
A typical measurement series is depicted in Figure~\ref{fig:Ex_3x4}. The top, middle, and bottom row of Figure~\ref{fig:Ex_3x4} show the spatial distributions of the time-averaged amplitudes, phase snapshots, and mean frequencies for different values of the external resistance. In the individual columns, the resistance values are: (a) $R=0.58 \rm \,k\Omega cm^2$, (b) $R=1.46 \rm \,k\Omega cm^2$, (c) $R=2.33 \rm \,k\Omega cm^2$, and (d) $R=3.65 \rm \,k\Omega cm^2$.
The amplitude distribution at the lowest global coupling (top row, column (a)) exhibits, over most of the electrode, a yellow-green color, corresponding to low values of the amplitude. Only at the right edge of the electrode does the red color indicate a small region with significantly higher amplitude. 
As the value of $R$ increases, this high-amplitude region increases in size ((b) - (d)), forming an L-shaped connected region in the upper right corner while the low-amplitude cluster around the lower left corner shrinks.
Additionally, with increasing resistance, the contrast between the two regions and thus the difference in amplitude becomes weaker.

The phase distributions shown in the middle row of Figure~\ref{fig:Ex_3x4} again reveal two distinct regions whose shapes are identical to those of the amplitude regions.
The red area with a phase close to $\pi$ corresponds to the low-amplitude region, and the green area corresponds to the high-amplitude region.
Here, the expansion of the high-amplitude region with resistance is seen even more clearly.
These plots indicate that the oscillators within each of the two regions are locked in phase, and the phase of the high-amplitude region is slightly ahead of that of the low-amplitude region.

This contrasts with the frequency distributions across the electrode, which are shown in the bottom row of Figure~\ref{fig:Ex_3x4}. 
The uniform color in each case signifies that, at any strength of the global coupling, the entire electrode oscillates at a uniform frequency.
The only difference between them is a slight increase in frequency with resistance.

\begin{figure}[htbp]
\centering
\includegraphics[width=0.8\textwidth]{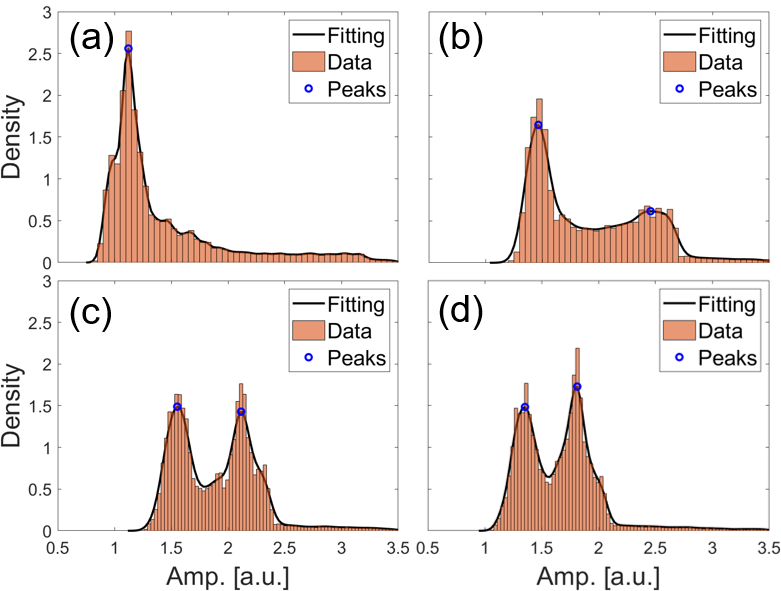}
\caption{
Histograms of experimentally measured time-averaged amplitudes at four different resistances, (a) $R=0.58 \rm \,k\Omega cm^2$, (b) $R=1.46 \rm \,k\Omega cm^2$, (c) $R=2.33 \rm \,k\Omega cm^2$, and (d) $R=3.65 \rm \,k\Omega cm^2$.
}
\label{fig:Ex_Histogram}
\end{figure}

\begin{figure}[htbp]
\centering
\includegraphics[width=0.6\textwidth]{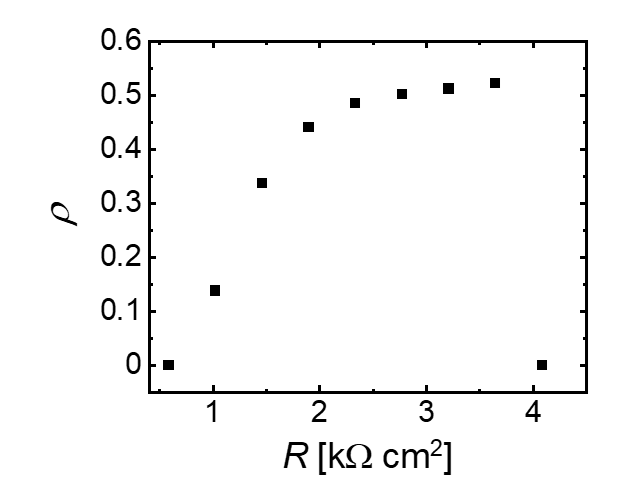}
\caption{
Relative size of the experimentally obtained high-amplitude region as a function of the resistance.
}
\label{fig:Ex_rho2}
\end{figure}

Figure~\ref{fig:Ex_Histogram} shows normalized histograms of the amplitude variations shown in the first row of Figure~\ref{fig:Ex_3x4}, obtained by kernel density estimation.
The orange bars, black curves, and blue points represent the data, the fitting, and the peak positions, respectively.
At the lowest resistance (a), the distribution is dominated by a pronounced peak at low amplitudes, accompanied by a long tail of high amplitudes.
With greater resistance, a second maximum emerges from this tail, the height of which increases with increasing resistance ((b)-(d)). The distributions thus become bimodal, reflecting the high- and low-amplitude regions in Figure~\ref{fig:Ex_3x4}. Corresponding histograms are obtained from the phase snapshots.
The relative sizes of the two regions were estimated by integrating the fit of the histograms around their maxima, taking into account $\pm 0.2$ amplitude units in the integration. 
The relative size of the high-amplitude clusters, $\rho$, is shown for nine different values of the external resistor in figure~\ref{fig:Ex_rho2}.
At low values of $R$, $\rho$ increases sharply to a value of $\approx 0.5$, from where it rises only slowly with resistance until, at $R\le 4\,\rm k\Omega cm^2$, the two regions merge into a single one again.  

In summary, the experiments show that for specific parameters, globally coupled electrochemical oscillators exhibit bimodal amplitude distributions accompanied by frequency synchronization.
Each mode manifests itself in a homogeneous phase distribution, while the two distributions exhibit a phase shift. 
The global coupling intensity $R$ controls the relative size of the amplitude modes.

\section{\label{sec:CGLEMFsec}Complex Ginzburg–Landau Equation with Mean-Field Coupling}
The above-described experiments on the anodic electrochemical etching of silicon revealed an intriguing kind of collective bimodal pattern in an oscillatory medium. These patterns are characterized by amplitude and phase distributions that exhibit two peaks with a particular variance. Remarkably, the local frequencies are identical (within our temporal resolution), i.e., the system exhibits complete frequency locking. A second intriguing feature is that the external resistance, which acts as a global coupling parameter (see Eq.~\eqref{eq:resitanceGC}), dictates the relative weight of the two modes.
These observations raise a natural question: What theoretical framework can capture both the emergence of these synchronized patterns and their organization in parameter space?

\subsection{\label{subsec:ModCGLE} Modified Complex Ginzburg-Landau Equation} 
To model the experiments, we employ a modified complex Ginzburg–Landau equation (CGLE), a standard amplitude equation for oscillatory reaction–diffusion systems~\cite{Haken.1984, GarciaMorales.2012}. The CGLE naturally captures the local oscillatory dynamics through Stuart–Landau terms, which describe the oscillations in the silicon system that arise through a Hopf bifurcation, as explained in Ref.~\cite{Schonleber.2014} and more recently justified mathematically in Ref.~\cite{Murakami.}.
We introduce a general mean-field coupling term to account for the experimentally observed control of the ratio of high- and low-amplitude regions through the external resistance. This accounts for the global electrical feedback (see Section~\ref{sec:ECSE}) acting equally on all oscillators. Since perfectly homogeneous conditions cannot be achieved in any electrochemical experiments, we consider spatial heterogeneity by assuming that the intrinsic frequencies $\omega(\vec{x})$ vary slightly across the electrode surface. Finally, spatial coupling through ionic and electronic transport must be considered because the silicon electrode is an extended medium; this effect is modeled by a diffusion term. Taken together, we end up with the heterogeneous, globally coupled CGLE:
\begin{equation}\label{eq:CGLE_het}
\begin{aligned}
    \partial_t  W(\vec{x},t) &=  W(\vec{x},t)(1-i\omega(\vec{x})) - (1+iC_2) \big| W(\vec{x},t) \big|^2 W(\vec{x},t) \\
    &\quad + K(1+iC_1)(\langle W(t)\rangle - W(\vec{x},t)) 
     + \epsilon(\partial^2_{x_1} +\partial^2_{x_2}) W(\vec{x},t),
\end{aligned}
\end{equation}
where the spatial mean is defined as:
\begin{equation}\label{eq:spatial mean}
\langle W(t)\rangle \coloneqq \frac{1}{L_1L_2}\int_0^{L1}\int_0^{L2}W(\vec{x},t)dx_1dx_2 ~.
\end{equation}
Here, $\vec{x}=(x_1,x_2)$ denotes the spatial coordinates on the rectangular domain $[0,L_1] \times [0,L_2]$. We can represent the complex variable $W$ in polar coordinates with radius $r$ and phase $\phi$:
\begin{equation}
\label{eq:Varible_polar_coordinates}
    W(\vec{x},t)=r(\vec{x},t)\text{e}^{\phi(\vec{x},t)}
\end{equation} Next, we discuss the model's parameters and their relationship to the experiment described above. The shear parameter of the individual oscillators, $C_2$, quantifies the amplitude–frequency interaction, whereas the $\omega(\vec{x})$ term specify the natural frequency of each oscillator at position $\vec{x}$. 
The global coupling parameters are $K$ (diagonal coupling) and $K C_1$ (off-diagonal coupling), where $K$ sets the coupling strength and $C_1$ modulates the phase-shift introduced by the coupling. 
Changes in the external resistance modify the potential drop (see equation~\eqref{eq:resitanceGC}) and, through this term, affect both the dissipative and reactive components of the effective coupling between oscillators. We therefore assume that the resistance $R$ strongly influences the phase-shift $C_1$. Since the microscopic form of the global coupling cannot be determined explicitly, we employ a generic complex coupling term that can account for a broad range of physically plausible global-coupling scenarios $K(1+iC_1)$. The diffusion strength is given by $\epsilon$. In summary, we can say that the Stuart-Landau oscillator qualitatively represents the local experimental oscillator. A system of Stuart-Landau oscillators interacting globally and through weak diffusion should provide qualitative insight into the emergence of the observed patterns.

\begin{figure*}
\centering
\includegraphics[width=0.9\textwidth]{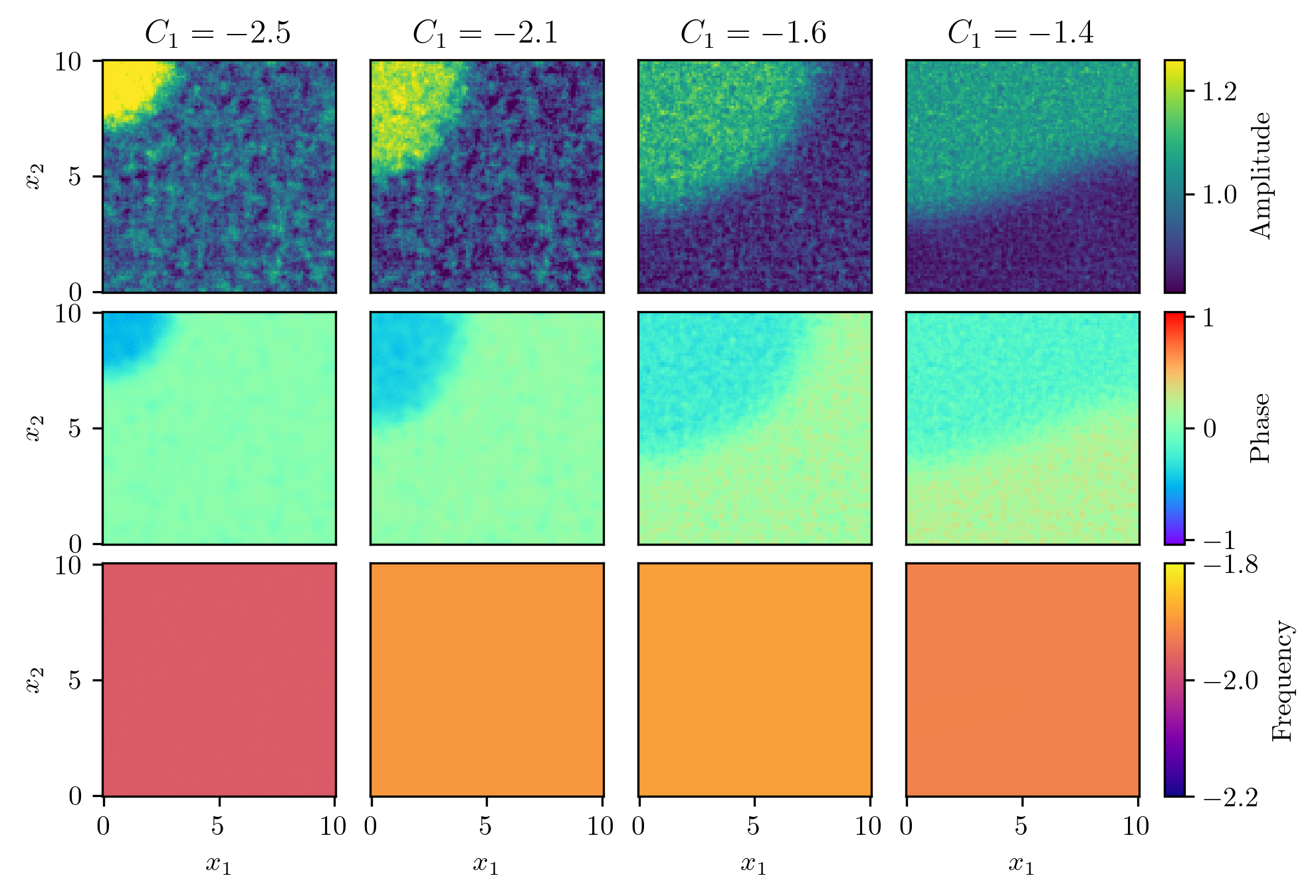}
\caption{\label{fig:Yuki_1_Sim} Numerically obtained snapshots of the amplitude (first row) and phase (second row) distributions, and frequency distribution (third row) for $K=1.15, C_2=2,\epsilon=0.02,\sigma_0=0.3$. The $C_1$ parameter is adiabatically changed, and the simulation results are plotted for four different values of $C_1$. First column $C_1=-2.5$, second column $C_1=-2.1$, third column $C_1=-1.6$, and last column $C_1=-1.4$.}
\end{figure*}\

We numerically solve Eqs.~\eqref{eq:CGLE_het} on a square domain ($L_1=L_2=10$) with Neumann boundary conditions, discretized on a $100\times 100$ uniform grid. To ensure reproducible frequency realizations, we generate $\omega(\vec{x})$ with Julia’s Mersenne Twister PRNG (seed \texttt{2024}). The intrinsic frequencies $\omega_k$ are drawn from a Gaussian distribution with mean $\mu_0$ and standard deviation $\sigma_0$; moving to a co-rotating frame with $\mu_0$ as its center shifts the distribution to zero. We use adiabatic continuation and integrate system~\eqref{eq:CGLE_het} for $T=1000$ time steps. Here, "adiabatic" means that we first run the system for $T$ time steps at $C_1 = C_1^0$. We then increase (decrease) $C_1$ to $C_1^0 + \Delta C_1$ ($C_1^0 - \Delta C_1$) and integrate for another T time steps using the final state of the previous run as the initial condition for the next one. The numerical integrations are performed using Julia's \texttt{DifferentialEquations.jl} ecosystem~\cite{Rackauckas.2017}. We use \texttt{Tsit5}~\cite{Tsitouras.2011}, a Runge-Kutta-based solver with adaptive timesteps, and save data points at $\Delta t = 0.1$. The parameters $K$ and $C_2$ are chosen to obtain solutions where phase-locked solutions with fixed amplitude exist~\cite{Nakagawa.1993}. 

Starting from $C_1=-1.75$, we vary the coupling parameter in increments of $\Delta C_1 = \pm 0.05$ until reaching $C_1=-1$ and $C_1=-3$, respectively. Figure~\ref{fig:Yuki_1_Sim} shows the numerical results of the spatial distribution of the amplitude (first row), phase (second row), and frequency (third row) for four different $C_1$-values. The amplitude, phase, and frequency are shown in snapshots. All results are recorded after a sufficient evolution time to ensure that transients have decayed.\ 

For $C_1=-2.5$, two distinct amplitude and phase domains are clearly visible: a high-amplitude region in the top-left corner and a low-amplitude region occupying the remaining area. The underlying eigenfrequency heterogeneity $g(\omega)$ is reflected in amplitude and phase patterns, yet the mean frequencies are spatially uniform. As $C_1$ increases, the high-amplitude region expands while the low-amplitude region shrinks, and the values of the amplitudes in the two regions approach each other. Furthermore, the frequencies for each $C_1$ value are uniform across the entire spatial domain. 
These plots show that the oscillators within each of the two regions are locked at similar phases, and the phase of the high-amplitude region is slightly ahead of that of the low-amplitude region (note that the frequency is defined negatively in this case).

\begin{figure}
\centering
\includegraphics[width=0.6\textwidth]{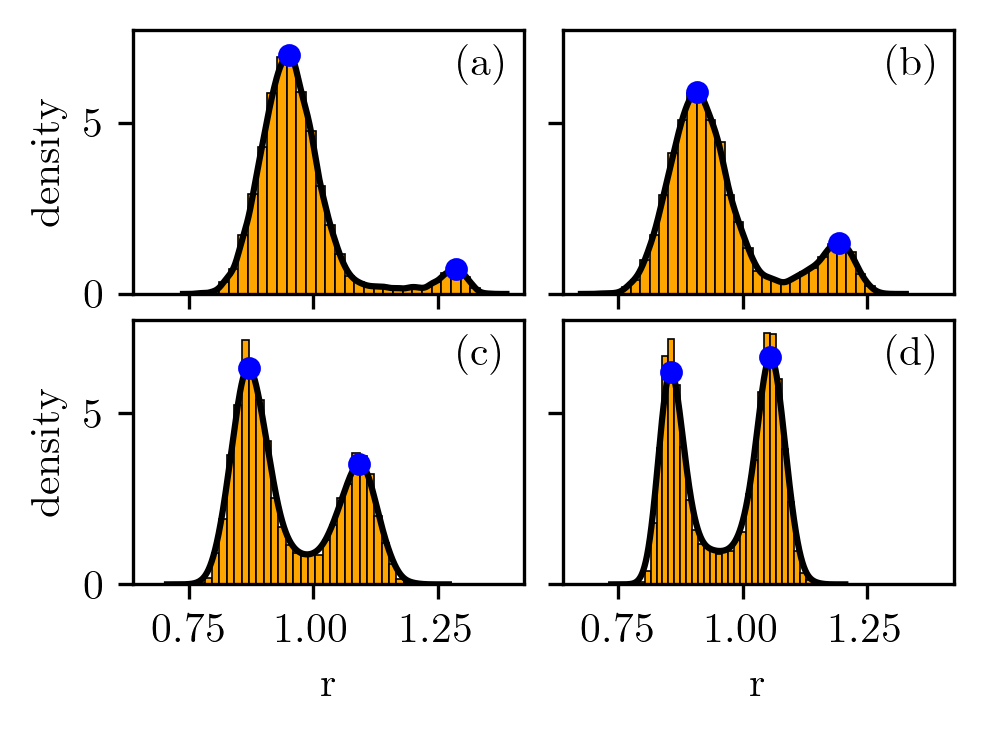}
\caption{\label{fig:Yuki_2_Sim} Histograms of the amplitudes $r$ for for different values of the global coupling parameter $C_1$. The fitting has been performed using the kernel density estimator. Parameter values: $K=1.15, C_2=2,\epsilon=0.02,\sigma_0=0.3$, (a) $C_1=-2.5$, (b) $C_1=-2.1$, (c) $C_1=-1.6$, and (d) $C_1=-1.4$.}
\end{figure}\ 

In Figure~\ref{fig:Yuki_2_Sim}, we plot the distribution of the amplitudes for the same $C_1$-values used in Figure~\ref {fig:Yuki_1_Sim}. The amplitudes can be extracted from the numerical data using Eq.~\eqref{eq:Varible_polar_coordinates}. From the data, it is clear that there are two peaks for all four values of $C_1$. Using the Kernel density estimate defined in equation~\eqref{eq:KDE}, we can locate the mean radius of each mode. The maxima of the distributions, shown as blue dots, are used to quantify the mode states via the ratio $\rho$ defined in Section~\ref{sec:ECSE}. From the four different histograms, one can clearly observe how oscillators transition from the low-amplitude mode to the high-amplitude mode as $C_1$ is increased.\

\begin{figure}
\centering
\includegraphics[width=0.6\textwidth]{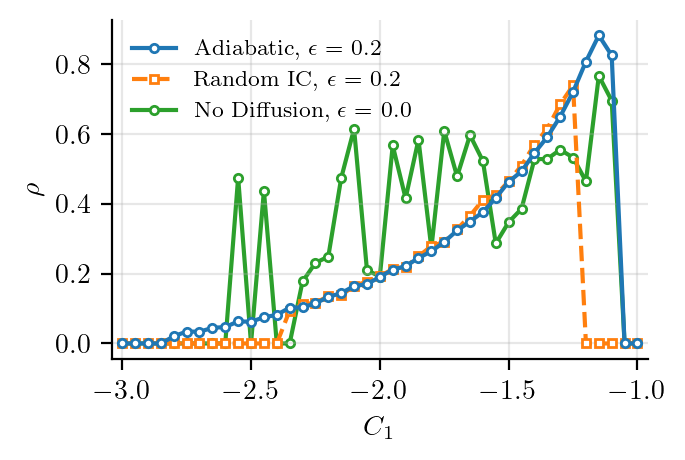}
\caption{\label{fig:Yuki_3_Sim} Ratio $\rho $ obtained for $K=1.15, C_2=2,\sigma_0=0.3$. $C_1$ is initiated at $-1.75$ and tuned to $C_1=-1$ and $C_1=-3$ by steps of $\pm0.05$. In blue, adiabatic continuation is performed with diffusion $\epsilon=0.02$. The same diffusion is used in orange, but each $C_1$ value starts from the same initial condition. The green line also starts from the same initial conditions without any diffusion.}
\end{figure}\ 

In Figure~\ref{fig:Yuki_3_Sim}, we plot the fraction of the high-amplitude mode $\rho$ as a function of $C_1$ for three different simulation schemes. The cluster ratio $\rho$ has been determined as explained for Figure~\ref{fig:Ex_rho2}. Since the experimental peaks are broader than the simulated ones, we restrict the counting to oscillators within $\pm 0.15$ of each peak maximum. 
A unimodal distribution is assigned to $\rho=0$. 
The blue open circles represent simulation results, where, for $C_1=-1.75$, a random initial condition was chosen. For the other values of $C_1$, the simulations were continued adiabatically. For the simulations shown in orange symbols, the same initial condition is chosen for each value of $C_1$, and the simulation shown in green is identical to orange, but without diffusion. First, consider the adiabatic continuation. Starting at $C_1=-3$, the system is in a synchronous solution. At $C_1\approx -2.8$, it jumps to a pattern of $\rho\approx0.06$. In this configuration, many oscillators are in the low-amplitude region, and few are in the high-amplitude region. Increasing $C_1$ leads to more oscillators transitioning to the high amplitude region until a critical ratio is reached at $C_1\approx -1.1$, at which the system transitions back to a unimodal distribution. For the continuation with a random initial condition, the $C_1$-interval where the bimodal pattern appears is narrower. For the no-diffusion series (green), the amplitude distribution is often not clearly bimodal; in those cases, the KDE peak–picking ansatz becomes ill-posed and the inferred $\rho$ fluctuates strongly. Therefore, these apparent “jumps” should be interpreted cautiously, as they largely reflect estimator instability rather than genuine rapid changes of the domain fraction.

\subsection{Comparison of experimental and numerical results}\ 

A comparison of Figures~\ref{fig:Ex_3x4}~--~\ref{fig:Ex_rho2} with Figures~\ref{fig:Yuki_1_Sim}~--~\ref{fig:Yuki_3_Sim} clearly shows that the experimental results and numerical simulations of eqs.~\eqref{eq:CGLE_het} exhibit the same collective pattern: a bimodal amplitude distribution with a spatially uniform frequency field where the high amplitude mode leads and the low amplitude mode trails in phase. Varying the control parameter tunes the relative size of the high-amplitude domain, with $\rho$ changing monotonically until the system eventually returns to a unimodal state. This evolution is reflected in the histograms, where two well-separated modes shift in relative weight before collapsing into one.

Despite this agreement, systematic differences remain. Experimentally, high-amplitude regions pin to edges and form anisotropic shapes, while in simulations, they nucleate as compact patches. Such discrepancies arise from inhomogeneities, gradients, and boundary effects absent in the model, which assumes randomly distributed parameters (frequencies) and Neumann boundary conditions. Quantitative amplitude, phase, and frequency deviations are also natural, since the SL oscillator provides only an idealized description of dynamics near the Hopf bifurcation~\cite{Haken.1984}.

Taken together, the comparison supports a common mechanism underlying both datasets: amplitude clustering with frequency locking. The model reproduces the core phenomenology, including the coexistence of two amplitude regions, a controllable and unique size ratio, and the merging of the two domains at large control.\ 

\section{\label{sec:Discussion}Discussion}

Given the convincing agreement between experiments and simulations outlined above, it is worthwhile to further analyze the characteristic features of the numerically obtained amplitude patterns in more detail.

The simulations confirm two remarkable properties of the patterns. First, although the systems are heterogeneous and the natural frequencies $\omega(\vec{x})$ are sampled from a Gaussian distribution, the measured frequencies $\Omega(\vec{x})$ in the bimodal pattern are spatially uniform. This indicates full frequency entrainment despite the imposed heterogeneity.
Second, under adiabatic continuation in the control parameter, the system does not exhibit parameter intervals with fixed size ratios $\rho$, as is typical for the globally coupled Stuart–Landau system  (see~\cite{Kemeth.2019}). Instead, the interface between the low- and high-amplitude domains shifts smoothly with the control parameter $C_1$, so that the size of the high-amplitude region, and hence $\rho$, changes \emph{continuously} with $C_1$. In other words, rather than selecting one of several discrete (“quantized”) cluster ratios, the modified heterogeneous CGLE exhibits a near-continuum of bimodal patterns parameterized by the interface position. The same behavior is observed using the same random initial condition at each parameter set. However, when diffusion is removed, the system jumps randomly between coexisting cluster solutions. In the following, we will discuss the origin of these two peculiar properties of the patterns. \

\subsection{\label{subsec:HChetSL}The Ghost of the Cluster Singularity in Globally Coupled Heterogeneous Stuart-Landau oscillators}\

We first address the question of the mechanism that leads to complete frequency entrainment of the heterogeneous medium with Gaussian-distributed frequencies $\omega(\vec{x})$ even at moderate coupling strengths. This differs from phase-only models, such as the Kuramoto model, in which complete synchronization is only observed when the coupling strength $K$ is at least equal to the frequency deviation between the fastest (or slowest) oscillator and the mean frequency~\cite{Strogatz.2000}. This suggests a crucial role of the amplitude degree of freedom. To understand the entrainment mechanism in amplitude oscillators, this section disregards the spatial extension, i.e., the diffusion term in Eq.~\eqref{eq:CGLE_het}, and studies an ensemble of only globally coupled Stuart-Landau oscillators. This complies with experimental conditions where the global coupling strength is much stronger than the diffusive coupling.   

We thus consider a system of $N$ globally coupled, heterogeneous Stuart–Landau oscillators that evolves according to
\begin{equation}
\label{eq:SLGC_het}
\begin{aligned}
\partial_t W_k &= W_k(1-i\omega_k) - (1+iC_2)\lvert W_k\rvert^2 W_k \\
  &+ K(1+iC_1)\bigl(\langle W\rangle - W_k\bigr), \\
\end{aligned}
\end{equation}
where $\omega_k$ are the intrinsic frequencies and $\langle W\rangle$ is the mean-field defined as:
\begin{equation}\label{eq:global mean}
\langle W\rangle = \frac{1}{N}\sum_{j=1}^NW_j~.
\end{equation}
In both experiments and simulations, we observed phase-locked solutions, which can be written in polar form  
\begin{equation}
\label{eq:phase-lock}
W_k(t) = r_k\,e^{i(\Omega t+\phi_k)}, \qquad
\langle W\rangle(t) = \mathfrak{R}\,e^{i(\Omega t+\psi)},
\end{equation}
with constant amplitudes $r_k$, phase offsets $\phi_k$, a collective amplitude $\mathfrak{R}$, and a common frequency $\Omega$. The order parameter $\mathfrak{R}$ quantifies population coherence. Exploiting rotational symmetry, we set $\psi=0$ so that all phases are measured relative to the mean. Under this choice, the introduced collective amplitude $\mathfrak{R}$ simplifies to 
\begin{equation}\label{eq:OP_het}
\mathfrak{R} \coloneqq \frac{1}{N}\sum_{k=1}^{N} r_k \cos\phi_k, \quad \text{with} \quad \sum_{k=1}^{N} r_k \sin\phi_k = 0,
\end{equation}
where $\mathfrak{R}$ measures the coherence, and the condition $\sum_kr_k\sin\phi_k=0$ ensures consistency with $\psi=0$. The mean frequency is $\Omega=\dot{\psi}$. The values of the coherence $\mathfrak{R}$ and the oscillator states $\{r_k, \phi_k\}$ is resolved self-consistently.
In the co-rotating frame at frequency $\Omega$, locked states satisfy $\dot r_k=0$ and $\dot\phi_k=0$. 
Substituting this ansatz into~\eqref{eq:SLGC_het} and separating real and imaginary parts yields algebraic balance equations for $\{r_k,\phi_k\}$:
\begin{equation}\label{eq:Re_Im_Part}
\begin{aligned}
    0 &= r_k(1-r_k^2)-Kr_k+K\mathfrak{R}(\cos\phi_k-C_1\sin\phi_k), \\
    \Omega &=-\omega_k -C_2r_k^2+\frac{K\mathfrak{R}}{r_k}(\sin\phi_k+C_1\cos\phi_k)-KC_1
\end{aligned}
\end{equation}

Global coupling enables frequency locking by pushing the dynamics of each oscillator towards those of the collective mean field. The frequency balance equation (second line of Eqs.~\eqref{eq:Re_Im_Part}) shows that the effective frequency of each oscillator consists of its intrinsic term $-\omega_k$, the coupling parameters $K C_1$, an amplitude-dependent shift $-C_2 r_k^2$, and a coupling-induced correction $\frac{K\mathfrak{R}}{r_k}(\sin\phi_k+C_1\cos\phi_k)$, whose strength is set jointly by the global coupling $K$ and the coherence $\mathfrak{R}$. The last two terms enable the oscillators to adjust their amplitudes off the unit circle, thereby adapting the frequency to a collective value $\Omega$. For locking to occur, the total coupling strength must be sufficiently large to overcome the heterogeneity in intrinsic frequencies $\omega_k$, ensuring the coupling-induced adjustments dominate individual dynamics and suppress phase drift. Unlike the case of heterogeneous phase oscillators, however, the coupling strength each oscillator $k$ experiences is modified by $1/r_k$, resulting in stronger coupling when amplitudes adjust according to the frequency deviations.\

At this point, it is useful to contrast the synchronized solutions of heterogeneous phase oscillators with those of amplitude–phase oscillators. In the Kuramoto model, the collective frequency of the locked state coincides with the mean of the natural frequency distribution. By contrast, in heterogeneous Stuart–Landau ensembles, the collective frequency $\Omega$ does not reduce to this mean value. Amplitude dynamics alter the balance, such that synchronization leads to a collective frequency distinct from the mean frequency of the uncoupled system. \

Returning to the frequency-locked amplitude mode, introduced in equations~\eqref{eq:phase-lock}, it would be desirable to study its emergence with linear stability analysis. However, this is not feasible in this case: the locked state is only implicitly defined through the self-consistency equations for $\Omega$ and $\mathfrak{R}$, which could not be obtained analytically. As a result, we focus on a qualitative and numerical analysis of how the system transitions from a unimodal to a bimodal amplitude distribution.\
Therefore, we analyze the fixed-point equations~\eqref{eq:Re_Im_Part}. By eliminating the phase variable $\phi_k$ via the identity $\sin^2\phi_k+\cos^2\phi_k=1$, we obtain an implicit equation for the amplitude $r$, writing $r_k \to r$ and $\phi_k \to \phi$ for clarity since we consider a single oscillator at fixed $\omega$:
\begin{equation}\label{eq:ssquared}
\begin{aligned}
F(r,\omega,K,C_1,C_2,\mathfrak{R},\Omega)=F(\vec{p})=& 
    -(1+C_1^2)K^2\mathfrak{R}^2+r^2\Bigl[1+(1+C_1^2)K^2+(1+C_2^2)r^4\\
    &+(\omega+\Omega)^2 +2K(-1+r^2+C_1C_2r^2+C_1(\omega+\Omega))\\
    &+2r^2(-1+C_2(\omega +\Omega))\Bigr]
    =0 \, .
\end{aligned}
\end{equation}
This is a third-order polynomial in $r^2$, and for a given $\omega$, it may have one or three real positive roots, corresponding to one or three positive values of $r$. The transition from one to three real solutions is indicative of a saddle-node bifurcation, where a pair of fixed points either emerge or annihilate.\

Due to the implicit dependence of $F(\vec{p})$ on $\mathfrak{R}$ and $\Omega$, we cannot continue the solution branches of the full system. However, we can numerically elucidate some properties of the transition from unimodal to bimodal states. Therefore, we fix 
$C_1 = -2$, $C_2 = 2$, draw the intrinsic frequencies as 
$\omega_k \sim \mathcal{N}(0,0.1)$, and set $N = 10^4$. 
We then numerically solve Eqs.~\eqref{eq:SLGC_het} for coupling strengths 
$K \in \{1.5, 1.3, 1.2, 1.0\}$. For each $K$, we determine the macroscopic parameters 
$\mathfrak{R}$ and $\Omega$ of the endstate of the simulation, and evaluate $F(\vec{p})$ for a range of $\omega$-values. 

\begin{figure*}[htb!]
\centering
\includegraphics[width=1\textwidth]{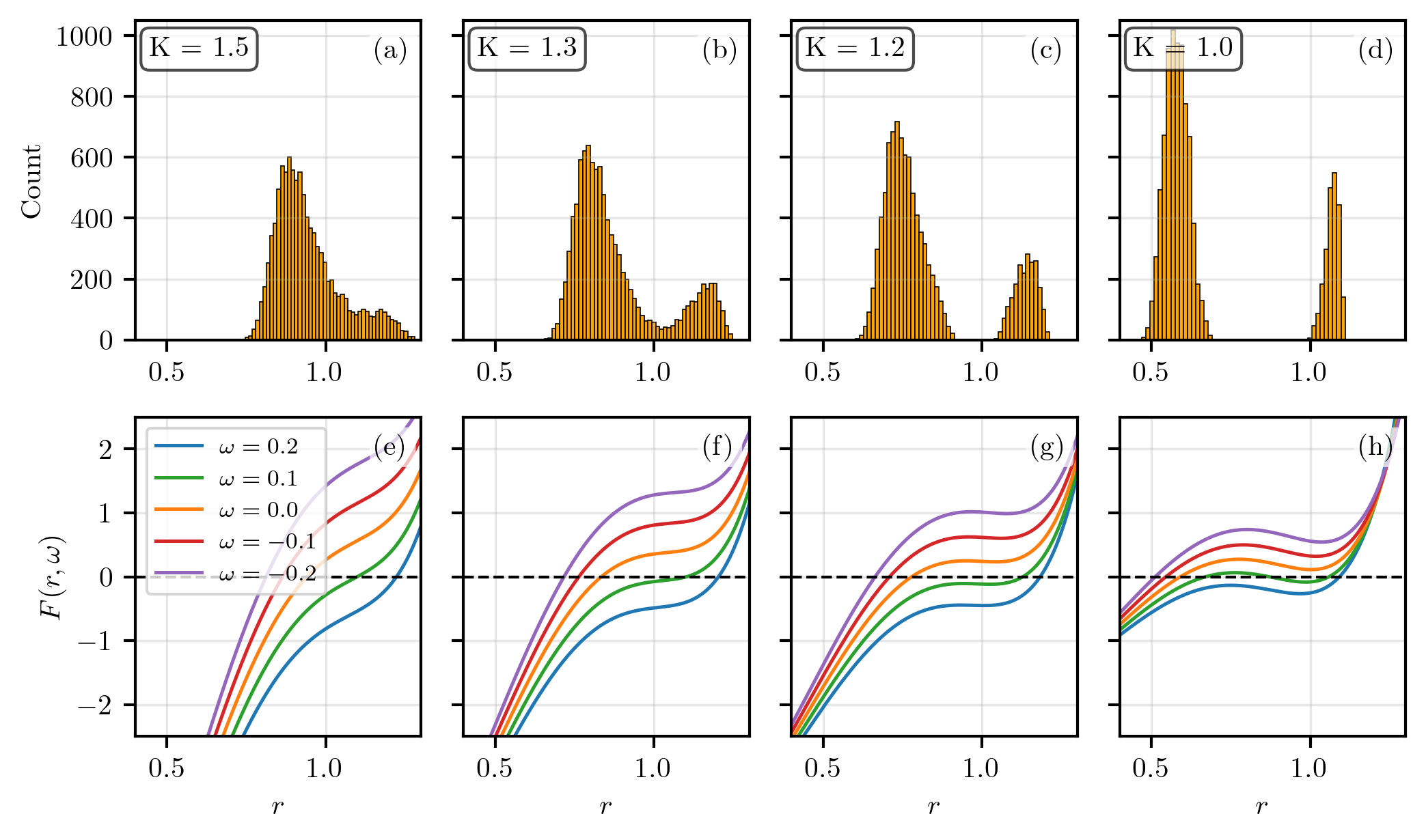}
\caption{\label{fig:Uni-Bimodal}Numerical simulation of system~\eqref{eq:SLGC_het} for $N=10000$. Top row: amplitude distributions for different values of $K$. Bottom row:  $F(\vec{p})$ for five different values of $\omega$ based on the parameters obtained from the simulations above. The same initial condition has been chosen for all simulations. The eigenfrequencies are pulled out of $\mathcal{N}(0,0.1)$ with the seed (MersenneTwister(2024). Other parameters are $ C_2 = 2$ and $ C_1 =- 2$. }
\end{figure*}

The resulting data are compiled in Table~\ref{tab:multistability}. Fig.~\ref{fig:Uni-Bimodal} depicts the amplitude distributions for decreasing $K$ (top row) together with fixed-point curves $F(r,\omega)$ for five representative $\omega$-values (bottom row).\

\begin{table}[htb!]
\centering
\renewcommand{\arraystretch}{1.2}
\begin{tabular}{cccc}
\toprule
\textbf{$K$} & \textbf{$\mathfrak{R}$} & \textbf{$\Omega$} & \textbf{Multistability Range} \\
\midrule
1.5 & 0.933 & -1.796 & monostable  \\
1.3 & 0.858 & -1.611 & monostable  \\
1.2 & 0.796 & -1.427 & $\omega\in${[}0.059, 0.063] \\
1.0 & 0.619 & -1.013 & $\omega \in${[}0.056, 0.13] \\
\bottomrule
\end{tabular}
\caption{Summary of numerical data; the collective amplitude $\mathfrak{R}$, the collective frequency $\Omega$ and the  investigation of multistability obtained for different coupling strengths $K$.}
\label{tab:multistability}
\end{table}

\

First, consider the amplitude distributions. At high coupling strength ($K=1.5$), the distribution is unimodal but skewed. 
The spread of amplitudes reflects that the system utilizes the amplitude degrees of freedom to compensate for frequency differences and achieve complete synchronization. 
As $K$ is lowered to $K=1.3$, the distribution becomes bimodal, but still extends over all amplitude values between the lowest and highest ones. 
This picture changes when $K$ is further reduced to $K=1.2$ and $K=1.0$. 
Here, the distributions consist of two separate parts with an unoccupied, or ``forbidden'' interval between them. 
It is also striking that at $K=1.0$, the two modes are considerably sharper than at $ K=1.2$.

Further insight into the transition from monomodal to bimodal distributions can be gained from the behavior of the roots of $F(r,\omega)$ for selected frequencies, as shown in the lower row of Fig.~\ref{fig:Uni-Bimodal}. 
In the first two cases at larger values of $K$, the $F(r,\omega)=0$ curves show a single root for every $\omega$. The stationary amplitudes vary only slightly for  $\omega=-0.1$ and $\omega=0$ but make a larger jump between $\omega=0$ and $\omega=0.1$, accounting for the broader spread at high amplitudes. 
This is particularly pronounced at $K=1.3$, where it leads to a bimodal shape of the amplitude histogram. 

At $K=1.2$, the $F(r,\omega)$ curves exhibit still a unique root for the values of $\omega$ shown, but in a narrow range of intrinsic frequencies $(\omega \in [0.059,0.063]) $, Eq.~\eqref{eq:ssquared} has three real roots. At $K=1.0$, the $\omega$-range of multiple roots broadens significantly, which is clearly visible in the $F(r,\omega=0.1)$ curve. The existence of this range implies that the system has undergone a series of saddle-node bifurcations. They occur when 
\[
 F(r,\omega) = 0 \quad \text{and} \quad \partial_r F(r,\omega) = 0
 \]
are fulfilled simultaneously. Simulations suggest that the outer one of the two new fixed points is stable while the inner one is unstable. For any amplitude distribution for which a frequency interval with three roots exists, any oscillator with an eigenfrequency in this interval may be in any of the two stable states. The occurrence of such an interval thus implies a high degree of multistability of the system.
Now, the origin of the forbidden amplitude zone is also clear. It emerges because all stationary solutions in this $r$-interval correspond to unstable states of the individual oscillator dynamics. Furthermore, the screwed and bimodal distributions at $K=1.5$ and $1.3$, respectively, can be interpreted as ``ghosts'' of the saddle-node bifurcations.\

Next, let us compare the dynamics of the heterogeneous SL oscillators described above with the known dynamics of identical SL oscillators. In the parameter range considered here, identical and globally coupled SL oscillators exhibit amplitude clusters, consisting of two internally synchronized groups that show a constant phase difference; each group has its own amplitude~\cite{Kemeth.2019, Kemeth.2020}. It therefore stands to reason that the bimodal amplitude distributions are the analog of 2-cluster solutions in heterogeneous oscillators. \

This assignment calls for further comparison: 
In the case of identical oscillators, the full $\mathbf{S}_N$ permutation symmetry of the dynamical system leads to a huge multistability~\cite{Golubitsky.2002} between different cluster states, particularly in the region where the homogeneous solution is unstable. Clearly, the heterogeneous system lacks a permutation symmetry. However, the bistable states of the individual oscillators discussed above allow for many different amplitude distributions to satisfy Eq.~\eqref{eq:ssquared}. We can thus view the coexistence of other distributions as the inheritance of an identical oscillator system and speculate that the multiplicity is greater the narrower the frequency distribution.\

Furthermore, in the homogeneous case, all the different 2-cluster states are born in saddle-node bifurcations which meet in a codimension-2 bifurcation, the cluster singularity~\cite{Kemeth.2019, Kemeth.2020}. The unfolding of the cluster singularity reveals an ordering of the 2-clusters, where the most unbalanced clusters are born at the borders of the 2-cluster region, and successively more balanced ones are formed as one moves toward the center of the region. This results in a characteristic fan-like structure of the most probable cluster solution, as shown in Fig. 2 of Ref.~\cite{Thome.2025}.
There are indications that relics of the cluster singularity also exist in our heterogeneous mean-field system. This concerns the saddle-node bifurcations that generate the distributions split into two parts. Moreover, we observed that the relative ratio of the amplitude modes $\rho$ tends to increase with increasing $C_1$, reminiscent of the fan-like structure. \rm

We will not elaborate further here on the 'ghosts' of the cluster singularity in the heterogeneous ensemble. Instead, to better align with our experimental result, we will next construct a bridge between the solutions of globally coupled oscillators and globally coupled oscillatory media in which diffusion is present but weak. More specifically, we will demonstrate that if diffusion is weak, it can be seen as a small perturbation that selects an energetically favorable specific size ratio or distribution $\rho$ out of the multiplicity of stationary stable solutions inherent to the $\mathbf{S}_N$ symmetry.

\subsection{\label{subsec:PatSelec} Pattern Selection Mechanism}
Above, we discussed common elements of the dynamics of identical and heterogeneous oscillators to shed light on the origin of the bimodal amplitude distributions observed in our experiments and with the heterogeneous CGLE with global coupling. Now, we will turn to the role of diffusion in shaping these patterns.\

We first assume the oscillators to be identical, and set $\omega(\vec{x})=\omega_0=0$. Furthermore, we consider the limit of small (non-zero) diffusion, $\epsilon\rightarrow0$, in which the spatial profile of the amplitudes and phases closely resembles a step function. This is justified by the numerical results, where diffusion layers between clusters remain narrow even for moderate values of $\epsilon$ (cf.~Fig.\ref{fig:Yuki_1_Sim}) and allows us to adopt an analytical approach inspired by Refs.~\cite{Dias.2003, Kemeth.2020}. To this end, we project the dynamics of the identical oscillators onto the center manifold of the Benjamin-Feir instability. This removes the neutral direction associated with phase invariance and reduces the dynamics of each oscillator to a scalar amplitude equation. The resulting one-component PDE reads:
\begin{equation}\label{eq:CMSLD}
\begin{aligned}
    \frac{\partial u}{\partial t} = \lambda u +& A(u^2 - \langle u^2 \rangle) + B(u^3 - \langle u^3 \rangle) + C \langle u^2 \rangle u + \epsilon \frac{\partial^2 u}{\partial x^2}.
\end{aligned}
\end{equation}

where $\lambda$ is a function of $K \text{ and } C_2$ and  $A, B \text{ and}, C $ are functions of $K, C_1 \text{ and }C_2$ of our full Eq.~\eqref{eq:CGLE_het}~\cite{Kemeth.2020}. The diffusion coefficient has not been subjected to the coordinate transformation, and we have added it to capture spatial structures. The variable $u(x,t)$ is thus a scalar function of space $x\in[0,L]$ and time, where we assume $x$ to be 1-dimensional. This scalar function can be treated as the amplitude of the Stuart–Landau oscillator. The global coupling is defined as: 
\begin{equation}\label{eq:u_mean}
    \langle u^n \rangle =\frac{1}{L} \int_0^Lu^n(x,t)dx.
\end{equation}

To study two-cluster configurations, we assume that $u(x)$ consists of two domains with piecewise constant values, separated by a sharp interface. The sharp interface is motivated by a small value of $\epsilon$. Specifically, we define: 
\begin{align}\label{eq:spatialDistr}
    u(x) =
\begin{cases}
u_1, & x \in [0, \rho L] \\
u_2, & x \in [\rho L, L]
\end{cases},
\end{align}
where, as above, $\rho \in[0,1]$ denotes the fraction of space occupied by the cluster with the high amplitude. By construction, system~\eqref{eq:CMSLD} is globally coupled through averages such as $\langle u^n \rangle$, which for a step function reduces to:
\begin{align}\label{eq:2Cluster_approach}
    \langle u^n \rangle = \rho u_1^n+(1-\rho)u_2^n,
\end{align}
Moreover, the center manifold reduction imposes the constraint $\langle u \rangle=0$~\cite{Kemeth.2020}, which yields:
\begin{align}\label{eq:u1u2}
    \rho u_1+(1-\rho)u_2=0 \Rightarrow u_2=\frac{-\rho}{1-\rho}u_1 .
\end{align}
This allows us to eliminate $u_2$ and express the entire configuration in terms of $u_1$ and $\rho$. Substituting into the original equation~\eqref{eq:CMSLD}, we obtain a closed PDE for $u_1$:  
\begin{equation}\label{eq:CMSLD1D}
\begin{aligned}
    \frac{\partial u_1}{\partial t} = \lambda u_1 +& A(u_1^2 - \langle u^2 \rangle) + B(u_1^3 - \langle u^3 \rangle) + C \langle u^2 \rangle u_1 + \epsilon \frac{\partial^2 u_1}{\partial x^2}.
\end{aligned}
\end{equation}
The above expression allows for two two-cluster fixed-point solutions~\cite{Kemeth.2020}, where each fixed-point solution expresses the amplitude of one of the clusters of the two-cluster solution. We note the two fixed points $u_1^{a,b}$. Notice that the sign of $u_1^a$ is always opposite to that of $u_2^a$. The expression of both fixed points is given by:
\begin{equation}
\begin{aligned}
u_1^{a,b} &= - \frac{(-1 + \rho) }{2 \left[ B + 3B(-1 + \rho)\rho - C (-1 + \rho)\rho \right]}\Bigl[\ A (-1 + 2\rho) \\
&\mp \sqrt{-4 B \lambda + A^2 (1 - 2\rho)^2 - 4 (3 B - C) \lambda (-1 + \rho)\rho} \Bigr]~,
\end{aligned}
\end{equation}
where $u_1^a$ corresponds to the stable branch~\cite{Kemeth.2020}.\ 

To understand how the system chooses a certain cluster configuration which is characterized by $\rho$, it seems natural to derive a Lyapunov functional $\mathcal{F}[u] $ such that the PDE of~\eqref{eq:CMSLD} corresponds to a gradient flow:
\begin{align}\label{eq:gradiantFlow}
    \frac{\partial u}{\partial t} = -\frac{\delta \mathcal{F}[u]}{\delta u}.
\end{align}
The functional derivatives reduce to ordinary derivatives for terms involving only local powers of $u(x)$. Using Eq.~\eqref{eq:2Cluster_approach}, which only holds for the case of small diffusion, we can argue that spatial means can be treated as scalars that depend on $\rho, \lambda, A, B \text{ and } C$. The diffusion term can be obtained using the Euler-Lagrange equation. The Lyapunov functional thus reads:
\begin{equation}
\begin{aligned}
\mathcal{F}[u] = - \int_0^L \biggl[ \frac{\lambda}{2} u^2 + \frac{A}{3} u^3 - A\langle u^2\rangle u + \frac{B}{4} u^4 - B\langle u^3\rangle 
    &+ \frac{C}{2} \langle u^2 \rangle u^2 - \frac{\epsilon}{2} (\partial_x u)^2 \biggl]dx
\end{aligned}
\end{equation}
As introduced above, the linear terms in $u$ vanish under the integral. We can thus rewrite the Lyapunov functional as:
\begin{align}\label{eq:Energy_functional}
\mathcal{F}[u] =-\frac{\lambda}{2}\langle u^2 \rangle - \frac{A}{3}\langle u^3\rangle  - \frac{B}{4}\langle u^4\rangle  - \frac{C}{2}\langle u^2\rangle^2 +\int_0^L\frac{\epsilon}{2}(\partial_x u)^2dx
\end{align}
For a small diffusion constant $\epsilon$, we can assume the contribution part of the integral to $\rightarrow 0 $. Using equations~\eqref{eq:u_mean} and~\eqref{eq:u1u2}, we find an analytical expression for $\mathcal{F}$:
\begin{equation}\label{eq:analyticalFormNoDiffusion}
\begin{aligned}
\mathcal{F}(\rho,u) = 
- \frac{\lambda L \rho u^2}{2(1 - \rho)} 
- \frac{A L \rho (1 - 2\rho) u^3}{3(1 - \rho)^2} 
- \frac{B L u^4}{4} \left( \rho + \frac{\rho^4}{(1 - \rho)^3} \right) 
- \frac{C L \rho^2 u^4}{2(1 - \rho)^2}.
\end{aligned}
\end{equation}
$u$ is either of the fixed points $u_1^{a,b}$ obtained from solving~\eqref{eq:CMSLD1D}. \newline \\
\begin{figure}
\centering
\includegraphics[width=0.7\textwidth]{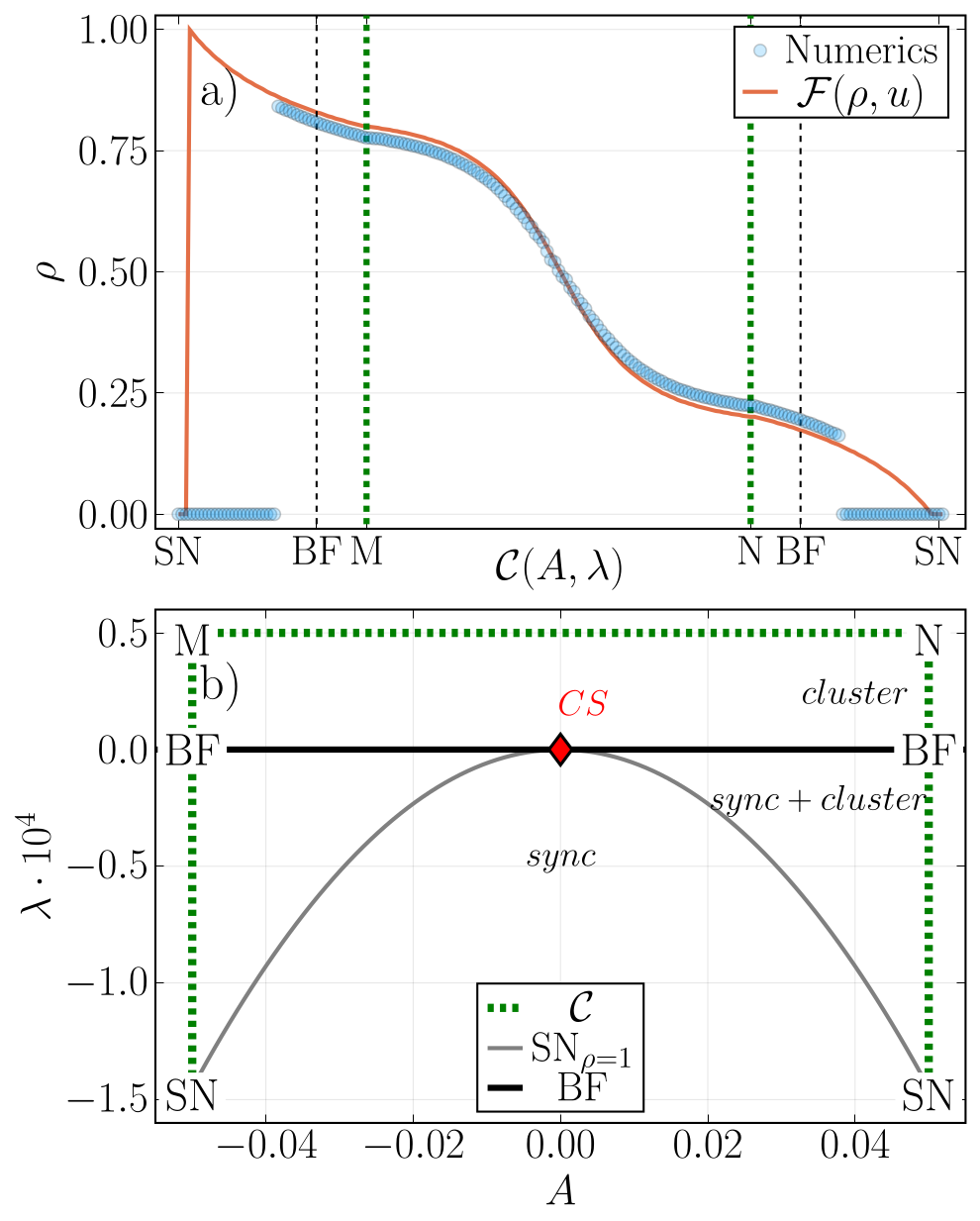}
\caption{a) Cluster selection according to numerical simulations with a diffusion of $\epsilon = 10^{-6}$, $B=\frac{-2}{2\sqrt{3}-3}$ and $C=-1$ in the blue dots and according to the minima of the Lyapunov functional $\mathcal{F}(\rho,u)$ (orange line) along the contour $\mathcal{C}(A,\lambda)$ plotted in part b). In (b), the Black line indicates the Benjamin-Feir instability, and the gray line indicates the saddle-node bifurcation of the solitary solutions. \textit{sync} indicates that only the synchronous solution is stable, \textit{cluster} indicates that 2-cluster solutions are stable and \textit{sync} + \textit{cluster} indicates that both are stable}
\label{fig:Cluster_distribution_A}
\end{figure}
To assess the predictive accuracy of the Lyapunov functional $\mathcal{F}(\rho,u)$ in capturing the selected cluster ratio $\rho$, we numerically solve equation~\eqref{eq:CMSLD} in one spatial dimension with Neumann boundary conditions for 800 oscillators and $L=10$. Simulations are initialized using a sigmoidal initial condition satisfying $ \langle u \rangle = 0 $, and parameters are chosen in the vicinity of the cluster singularity (CS). Following Ref.~\cite{Kemeth.2020}, we fix
\begin{equation}
    B=\frac{-2}{2\sqrt{3}-3}, \quad C=-1,
\end{equation}
and set the diffusion constant to the small value $\epsilon = 10^{-6}$.
To systematically explore the cluster selection behavior near the CS, we define a piecewise-continuous path $\mathcal{C}(A,\lambda)$ in parameter space, composed of three segments which are represented in part (b) of Fig.~\ref{fig:Cluster_distribution_A}:
\begin{enumerate}
    \item A vertical path for fixed $A = -0.5$ starting on the saddle node bifurcation that creates the solitary solution (denoted as SN in Fig.~\ref{fig:Cluster_distribution_A}, passing the Benjamin–Feir instability (BF) and then continuing to the point M.
    \item A horizontal path varying $A$ for fixed $\lambda = 0.00005$ starting at the point M and going to the point N.
    \item A second vertical path varying $\lambda$ for fixed $A = 0.5$, starting at N, passing the BF stability boundary, and finishing at the SN.
\end{enumerate}

The path $\mathcal{C}$ begins and ends at saddle-node bifurcations of solitary states, whose location in parameter space can easily be derived from Ref.~\cite{Kemeth.2020}:
\begin{equation}
    \lambda_{SNs}=(3-2\sqrt{3})\frac{A^2}{3}~.
\end{equation}
The contour $\mathcal{C}$ is designed to wind around the CS so that solutions exist along the path for all $\rho \in [0,1]$. The Benjamin–Feir (BF) instability marks the boundary of stability for the synchronous solution: for $\lambda > 0$, only the cluster solutions are stable, whereas for $\lambda < 0$ and above the SN line, both the synchronous solution and cluster solutions are stable. Along the entire contour, cluster solutions are multistable. 

Along $\mathcal{C}(A,\lambda)$, we compute the selected cluster ratio $\rho$ from simulations and compare it to the minimizer of the Lyapunov functional $\mathcal{F}(\rho,u)$ evaluated for the same parameters at the fixed point $u_1^a$. This allows us to directly test the predictive capability of $\mathcal{F}$ across different regimes of the $(A,\lambda)$ plane, including near the CS where dynamics are expected to be most sensitive. As apparent from Figure~\ref{fig:Cluster_distribution_A}, the numerical data (blue circles) align closely with the predictions from the Lyapunov functional (red curve), except in narrow regions near sharp transitions. 
This result highlights the organizing role of the CS, which is strictly present only in systems of identical, globally coupled oscillators, in mediating transitions between qualitatively distinct macroscopic states. It highlights the role of diffusion in cluster selection within oscillatory media.\

In addition to cluster selection, diffusion also appears to define a critical domain size beyond which clustered areas do not exist:  in both experiments and simulations, the ratio of high- and low-amplitude domains could only be tuned within an intermediate region, see Figures~\ref{fig:Ex_rho2} and~\ref{fig:Yuki_3_Sim}. Close to $\rho=0$ and $\rho=1$, the system spontaneously returned to a unimodal distribution. Such a threshold is absent in discrete systems without diffusion. The synchronous solution coexists with cluster states at ratios close to 0 or 1, while diffusion naturally biases the system toward homogeneity. Two mechanisms may account for this transition: (i) the finite width of the diffusion layer, set by the ratio of the characteristic time scales of diffusion and reaction, destroys clusters smaller than this scale; and (ii) in two-dimensional media, the larger diffusion flux across curved fronts requires a critical nucleus size, below which domains collapse and the system reverts to a homogeneous profile~\cite{Haken.1994, Haken.1996}.\

At least phenomenologically, there appears to be a certain similarity between the amplitude cluster patterns described here and stationary domain patterns in bistable media with negative global coupling or feedback. 
In those systems, individual elements are bistable, and within a certain parameter range, a global constraint (e.g., a fixed mean field or imposed current) prevents the complete expansion of one state at the expense of the other, thereby stabilizing domain coexistence. 
By adjusting a parameter that impacts the global coupling, the size ratio of the two domains can usually be tuned, similar to what we describe above for the relative size of the cluster domains. 
Examples of such bistable systems include lithium-ion batteries, where at intermediate charging states, charged and discharged nanoparticles coexist, rather than a state composed of particles with an identical partial charging state~\cite{Dreyer.2011}. 
Other examples include hot-wire barretters, which can serve as voltage stabilizers~\cite{Barelko.1981, Haken.1994} or the electrochemical oxidation of CO on Pt under galvanostatic conditions~\cite{morschl.2008, bozdech.2018, Salman.2020}. A notable difference is that all individual, uncoupled oscillators have a unique amplitude, with a value that differs from the amplitude values of both domains. This implies that the existence of different states is coupling-induced, whereas in bistable systems, the states are determined by the dynamics of the individual elements, with the global coupling being responsible for their stable coexistence. In the oscillatory system, cluster or domain formation requires a finite value of $C_1$ in  Eq.~\eqref{eq:CGLE_het}. The variable $C_1$ translates to a cross-coupling of two variables in a physical system. In the one-component normal form~\eqref{eq:CMSLD}, the cross-coupling manifests itself in the nonlinear coupling terms. In contrast, the global coupling in the one-component bistable system can be linear, the necessary nonlinearity for coexisting domains stemming from the dynamics of the individual elements. It thus seems that in the case of amplitude clusters, the complexity in the evolution equations is shifted from the individual element to the coupling. In what ways these two methods for generating domain coexistence can be mapped to each other remains an interesting open question.

\section{\label{sec:Conclusion}Conclusion \& Outlook}
We have combined experimental and theoretical studies of bimodal patterns in a globally coupled oscillatory medium. In our experiments on anodic silicon dissolution, we observed two coexisting amplitude domains, with the instantaneous frequency remaining spatially uniform. The relative areas of the amplitude domains could be tuned smoothly until the minority phase collapsed, and synchrony was restored. We reproduced these dynamics with a heterogeneous, mean–field–coupled CGLE. Further insights into the role of heterogeneity, diffusion, and the mechanistic origin of the amplitude patterns were obtained with the help of subsystems.

Neglecting diffusion and studying heterogeneous mean-field coupled SL oscillators revealed that the frequency heterogeneities are absorbed in the amplitude degree of freedom, which allows complete synchronization~\footnote{The same qualitative effect occurs if heterogeneity is introduced via parameters other than $\omega_k$, such as the shear parameter $C_{2,k}$.}. The transition of the amplitude distributions from unimodal to bimodal could be related to the occurrence of saddle-node bifurcations in oscillators with a certain interval of heterogeneous frequencies, leading to a high degree of multistability, which can be easily verified numerically. A comparison with the dynamics of homogeneous mean-field coupled oscillators revealed that the organizing structure imposed by the cluster singularity on the two–cluster states in homogeneous Stuart–Landau ensembles persists in the presence of heterogeneity and diffusion. In other words, the ghost of the cluster singularity, which is a genuine property of systems with $\mathbf{S}_N$-symmetry, is still felt in the presence of small heterogeneities and diffusion.

Neglecting heterogeneities, on the other hand, we can show that diffusion selects a unique ratio of the two clusters from the many coexisting cluster states. Using a center-manifold and sharp-interface reduction, we derived an explicit Lyapunov functional whose minima predicted the selected cluster ratio. These predictions are in quantitative agreement with full simulations.\ 
 
In short, our results show that synchronization in amplitude oscillators is facilitated by absorbing the heterogeneous eigendynamics in amplitude variations, and diffusion in mean-field coupled oscillatory media drives the highly multistable system to a unique state, in accordance with recent works~\cite{Millan.2025}. Finally, we compared the mechanisms generating amplitude clusters in oscillatory media with the formation of two groups or domains in bistable systems with negative global coupling, which seem to share many properties.\

This work raises several open questions. On the experimental side, further investigations are needed to clarify why the external resistance in the electrochemical system effectively acts as a cross-coupling term. Another question concerns stochastic perturbations, which are present in any experiment and may influence the stability and selection of patterns, but were not considered in this study. Moreover, it would be important to analyze the transition into incoherence for the heterogeneous system of globally coupled SL oscillators~\eqref{eq:SLGC_het}. Numerical evidence suggests that there is a discontinuous transition off frequency synchronization as $K$ is decreased past a critical parameter $K_c$. In this context, it is interesting to investigate how far it is related to more recently reported synchronization transitions, such as explosive~\cite{GomezGardenes.2011} or extreme synchronization~\cite{Lee.2025}.

\bibliographystyle{unsrtnat}
\bibliography{ClusterPartitioning}  






\end{document}